# Cryogenic Performance Evaluation of Commercial SP4T Microelectromechanical Switch for Quantum Computing Applications

*Yong-Bok Lee[1], Connor Devitt[2], Xu Zhu[3], Nicholas Yost[3], Yabei Gu[3], Sunil A. Bhave[2]\**

[1]Department of Electronics & Computer Engineering, Chonnam National University, 77 Yongbong-ro, Buk-gu, Gwangju 61186, Korea

[2]OxideMEMS Lab, Elmore Family School of Electrical and Computer Engineering, Purdue University, West Lafayette, IN 47907 USA

[3]Menlo Microsystems, Inc., Albany, NY 12203 USA.

\*Correspondence to: bhave@purdue.edu

**ABSTRACT**

Superconducting quantum computers have emerged as a leading platform for next-generation computing, offering exceptional scalability and unprecedented computational speeds. However, scaling these systems to millions of qubits for practical applications poses substantial challenges, particularly due to interconnect bottlenecks. To address this challenge, extensive research has focused on developing cryogenic multiplexers that enable minimal wiring between room-temperature electronics and quantum processors. This paper investigates the viability of commercial microelectromechanical system (MEMS) switches for cryogenic multiplexers in large-scale quantum computing systems. DC and RF characteristics of the MEMS switches are evaluated at cryogenic temperatures (< 10 K) through finite element simulations and experimental measurements. Our results demonstrate that MEMS switches exhibit improved on-resistance, lower operating voltage, and superior RF performance at cryogenic temperatures, with reliable operation over 100 million cycles. Furthermore, stable single-pole four-throw (SP4T) switching and logical operations, including NAND and NOR gates, are demonstrated at cryogenic temperatures, validating their potential for quantum computing. These results underscore the promise of MEMS switches in realizing large-scale quantum computing systems.

**Introduction**

Quantum computers have garnered significant attention in recent years due to their potential to solve certain problems with remarkable speed and efficiency by leveraging quantum phenomena such as entanglement, superposition, and interference[1]. Unlike classical computers, which rely on binary bits as the fundamental unit of computation, quantum computers utilize quantum bits, or qubits, capable of representing multiple states simultaneously. The implementation of qubits can be realized through various approaches, including superconducting circuits[2], trapped ions[3], photon polarization[4], and electron spins[5]. Among these, superconducting quantum computers have emerged as the leading platform due to their high scalability, rapid computational speed, and compatibility with semiconductor technologies[6]. Superconducting quantum computers generally consist of a quantum processor and control/read-out electronics as shown in Fig. 1a. Since quantum processors are highly sensitive to external environmental factors such as heat and electromagnetic fields, they are typically located at the 10 millikelvin (mK) base temperature stage of a dilution refrigerator, and connected to room-temperature control and read-out electronics through wide-band cables[7]. While this architecture is effective for experimental verification with a small number of qubits, scaling to the millions of qubits required for practical quantum applications[8,9] presents significant interconnect challenges due to the limited mK temperature stage area and cooling power of dilution refrigerator.

To address these challenges, new architectures utilizing cryogenic multiplexers have been actively researched to prepare the large-scale quantum computing system[10-12] (Fig. 1b). In this architecture, DC and RF signals from the room-temperature electronics are transmitted to the mK temperature stage, where the quantum processor is located, using only a small number of cables. At the mK temperature stage, these signals are distributed to multiple qubits or quantum information is collected from multiple qubits through the multiplexers. Since it enables effective communication between the room-temperature electronics and the quantum processor at 10 mK using a minimal number of wires, it is considered a next-generation interconnect architecture for realizing large-scale quantum computers with more than 1 million qubits. However, in order to implement reliable large-scale quantum computing with multiplexers, the switching device for the multiplexer must operate stably without performance degradation even at the cryogenic temperature of 10 mK, and the power consumption should be lower than the cooling budget of the dilution refrigerator, approximately 20 μW[13]. Furthermore, the device should exhibit low insertion loss (less than 0.5 dB), high isolation (greater than 30 dB), and fast switching capabilities within the 4 to 8 GHz qubit frequency range[10]. Various research groups are developing switching devices based on CMOS[10], HEMTs (high electron mobility transistors)[12], nanowires[14], and Josephson bifurcation amplifier[15] to implement cryogenic multiplexers. However, current studies face challenges in achieving stable operation, low power consumption, low insertion loss, and high isolation at cryogenic temperatures.

Microelectromechanical system (MEMS) switches have recently emerged as a promising alternative for implementing large-scale quantum computing. The MEMS switches offer near-zero static power consumption[16, 17], a high on/off ratio[18], and low sub-threshold swing (SS)[19, 20] due to their mechanical operating mechanism. Furthermore, with precise mechanical and material engineering, MEMS switches can operate stably even at the cryogenic temperature environments[21, 22], However, large-scale quantum computing systems with more than 1 million qubits would require a substantial number of reliable MEMS switches. Therefore, the use of commercially manufactured MEMS switches, which ensure high yield and consistent quality, would be ideal solution for future quantum computing systems.

In this paper, the DC and RF characteristics of commercial RF MEMS switches are evaluated at sub-10 K temperatures, and their applicability to quantum computing systems is explored. To investigate the cryogenic characteristics of the commercial MEMS switches, finite element method (FEM) simulations and experimental measurements of their DC and RF performance at cryogenic temperature are conducted. The results revealed that the MEMS switches exhibit a lower operating voltage, lower on-resistance, and improved RF performance compared to their room temperature results. Additionally, the MEMS switches exhibit high reliability over 100 million cycles and stable SP4T operation characteristics, showcasing their potential role in quantum computing systems. Furthermore, it is demonstrated that logical operations, such as NAND and NOR gates can be performed using the MEMS switches by leveraging its unique operating mechanisms.

**Results**

**Commercial RF MEMS Switch for Large-Scale Quantum Computing Systems**

Figure 1c-d presents an optical image and block diagram of the single-pole four-throw (SP4T) MEMS device developed by Menlo Microsystems, inc. The device is configured as a SP4T device with center input, which can be routed to one of the four outputs by applying a voltage to the corresponding gate electrode. Each individual MEMS switch features a cantilever structure[23]. In the absence of an operating voltage at the gate, the cantilever beam remains physically separated from the output signal line. When a gate voltage is applied, the cantilever beam is mechanically deflected and contacts the output signal line, enabling signal transmission. This mechanical operation results in ultra-low insertion loss, high isolation, and superior linearity. Furthermore, the MEMS switch is hermetically sealed through wafer level chip scale package (WLCSP) technology, enhancing its operational reliability. Therefore, the developed MEMS switch is widely utilized in various applications, including wireless communication, defense/aerospace, and automatic test equipment (ATE) and test instrumentation. In this study, the MEMS switch's properties at cryogenic temperatures are characterized to evaluate its potential applicability to quantum computing systems.

**DC and RF Performances Evaluation at Cryogenic Temperature**

First, the electrical characteristics of the MEMS switch, a fundamental component of the SP4T device shown in Fig. 1d, were assessed through FEM simulation. Air gap changes due to deflection significantly impact the performance of the MEMS switch[24], so deflection as a function of temperature was also simulated (Fig. 2a). The cantilever switch exhibited minimal z-direction displacement with temperature changes, as it can freely expand in length, unlike the fixed-fixed beam structure. The maximum displacement of approximately 60 nm occurred when the MEMS switch was exposed to the cryogenic temperatures. This result indicates that the MEMS switch can maintain consistent operating characteristics even at cryogenic temperatures since the change in air gap was insignificant. Indeed, as shown in Fig. 2b, the operating voltage decreased slightly by approximately 3.5% at 0 K compared to the room-temperature result. The robust cryogenic performance of the MEMS switch, predicted through FEM simulations, was further confirmed experimentally. All cryogenic experiments were conducted at approximately 5.8 K. Figure 2c presents the pull-in voltages of 16 MEMS switches measured at room temperature (left), at cryogenic temperature (middle), and after returning to room temperature (right). In alignment with the simulation results, the operating voltage at 5.8 K exhibited a slight reduction of approximately 3.1% due to the small decrease in the air gap distance. Notably, this result is significant compared to many conventional MEMS switches, which experienced substantial shifts in pull-in voltage at cryogenic temperatures[25-29]. Furthermore, the pull-in voltages returned to its original value after the MEMS devices were brought back to room temperature. Figure 2d shows the measured on-resistance (the resistance between the two interfaces during mechanical contact) at both room temperature and cryogenic temperature. The on-resistance was decreased by approximately 15.3% at cryogenic temperatures. The decreased on-resistance is attributed to the diminished phonon scattering in metals at lower temperatures[30]. Furthermore, the decreased on-resistances were fully recovered to their original values after returning to room temperature. The RF characteristics, including insertion loss and isolation, were also evaluated at the cryogenic temperature. Figure 2e shows the measured insertion loss at both room temperature and cryogenic temperature. The RF measurement was conducted between output_1 and output_3 (or output_2 and output_4) of the SP4T device as shown in Fig. 1d. At both temperatures, the insertion loss remains below 0.5 dB within the qubit frequency range of 4–8 GHz due to the low on-resistance of the MEMS switch. Notably, the values measured at cryogenic temperature were slightly improved compared to those at room temperature due to the effect of lower resistivity at low temperatures. The isolation characteristics were also measured as shown in Fig. 2f. The MEMS switch exhibited excellent isolation performance, exceeding 35 dB, due to the presence of an air gap in the off state. Importantly, the MEMS switch maintained exceptional isolation performance at cryogenic temperatures, as the airgap distance remained consistent even at the cryogenic temperature. In conclusion, these simulation and measured results support that the MEMS switch maintains its outstanding performance without degradation at the cryogenic temperatures, making it a promising candidate for quantum computing applications.

**Dynamic Response of the MEMS Switch**

The transient response of the MEMS switch was also measured through a voltage divider circuit to evaluate the switching speed and dynamic characteristics as shown in Fig. 3. The voltage divider circuit is shown in Supplementary Fig. 1a. This circuit consists of a MEMS switch connected in series with a load resistor. When the MEMS switch is in the off state, the voltage from the power supply is entirely applied across the MEMS switch due to its high off-resistance. In contrast, when the MEMS switch is turned-on by applying a gate pulse, the voltage from the power supply is distributed to the load resistor. Figure 3a-b shows the transient response when 10 kHz gate pulses were applied to the MEMS switch. Due to mechanical movement of the cantilever, contact occurred approximately 2.7 μs after the gate pulse was applied to the device, which is the switching speed. However, at cryogenic temperature, the transient response exhibits a different behavior, as shown in Fig. 3c-d. While the gate voltage is off, severe oscillations in the output voltage were observed. This behavior can be attributed to the WLCSP packaging of the MEMS switch. At cryogenic temperatures, the gas inside the package undergoes a phase transition, creating a quasi-vacuum environment. The absence of air damping in the cryogenic environment can lead to a pronounced bouncing phenomenon. Supplementary Fig. 1b shows the dynamic response over an extended period, illustrating that the bouncing persisted for approximately 150 μs. To determine whether the bouncing phenomenon is temperature-dependent, the dynamic response was measured as the temperature decreased from 100 K to 10 K. As shown in Supplementary Fig. 2, the bouncing phenomenon began near the boiling points of oxygen and nitrogen. These measurement results provide clear evidence that the bouncing effect was caused by a phase transition of the gases inside the packaging. This phenomenon at the cryogenic temperature can present significant challenge to the stable dynamic operation of the MEMS switch in the cryogenic application. To address this issue, an engineered waveform employing a dual-pulse approach is introduced, as shown in Fig. 3e–f, to ensure stable operation even at cryogenic temperatures. This method tailors the actuation waveforms to minimize the velocity of the cantilever upon contacting the bottom electrode, effectively suppressing bouncing[31-33]. The introduced engineered waveform has four distinct regions, departing from the typical square pulse in Fig. 3b. A detailed illustration of the width and period of the engineered waveform is shown in Supplementary Fig. 3. In the red region (90 V for 2 μs) as shown in Fig. 3f, a voltage higher than the pull-in voltage was applied to provide sufficient momentum for the switch to contact the bottom electrode. During the subsequent orange region (55 V for 1 μs), where a voltage lower than the pull-in voltage was applied, the switch contacts the bottom electrode with near-zero velocity. Finally, by immediately applying the holding voltage (90 V), the oscillation during the contact can be reduced. The same strategy was also used during the release process to reduce the bouncing of the cantilever. In this step, the blue region (80 V for 2 μs), where a voltage higher than the pull-in voltage is applied, immediately follows the green region (0 V for 1 μs). This sequence can bring the velocity of the cantilever to zero, enabling smooth detachment without the occurrence of bouncing. While this engineered waveform slightly increased the switching speed (approximately 3.3 μs), it can ensure stable dynamic operation even at the cryogenic temperature. Given that the MEMS switch operates at 10 kHz, the power consumption of a single MEMS switch at this frequency is calculated as follows: $P = \frac{1}{2}CV^2 f = \frac{1}{2} \times 12\text{fF} \times (90\text{V})^2 \times 10\text{kHz} = 0.607\ \mu\text{W}$. This ultra-low power consumption makes it highly suitable for quantum computing applications.

**Lifetime and logical operation of the MEMS switch**

Using the engineered waveform, reliability tests are conducted at cryogenic temperature. Figure 4 presents the measurement results of the dynamic response after $10^3$, $10^6$, $10^7$, and $10^8$ cycles. This is consistent with previous study showing that cryogenic temperatures under the boiling temperature of the oxygen enhance the reliability of the MEMS switch by suppressing native oxide formation at the contact interface[21]. Furthermore, the cantilever structure of the MEMS switch experienced minimal stress, as it can freely deform as the temperature decreases. This structural advantage indicates that the MEMS switch can maintain operational stability even at cryogenic temperatures. Consequently, the MEMS device demonstrated stable operation with no observable performance degradation after exceeding 100 million cycles. The overlapping graphs corresponding to these measurements are provided in Supplementary Fig. 4. These results show that the operating characteristics of the MEMS switch remain constant for 1 million cycles. Next, the operational stability of the SP4T device composed of MEMS switches under cryogenic temperature is examined. The SP4T MEMS device comprises one input and four outputs, allowing the input signal to be routed to a desired output by applying a gate voltage to the corresponding gate electrode. Figures 5a–c show the measurement results for routing the input signal to *output_1* when the operating voltage is applied to the *gate_1* electrode. Without the gate voltage, the input signal was not transmitted to the *output_1* electrode. However, upon applying the gate voltage, the input signal was successfully measured at the output_1 electrode. Similarly, the input signal was routed to *output_2* when the operating voltage was applied to the *gate_2* electrode (Fig. 5d-f). The measurement results with the signal routed to *output_3* and *output_4* are also provided in Supplementary Fig. 5. These results indicate that the SP4T MEMS device reliably routes the input signals to the desired output electrode even at cryogenic temperatures, demonstrating its potential for quantum applications. Additionally, NAND and NOR logic operations using the MEMS switches are demonstrated at the cryogenic temperature. Figures 6a–c show the measurement results of the NAND operation at approximately 5.8 K utilizing the interconnect structure of the SP4T device. The NAND operation was implemented by configuring a circuit where one load resistor and two switches are connected in series, as illustrated in Fig. 6a. In this setup, the two inputs correspond to the gate voltages of the two switches, and the output is the voltage across the MEMS switches connected in series. Utilizing the design of the SP4T MEMS device, which features four ports connected to a central input, the NAND operation can be performed as shown in Fig. 6b. Figure 6c shows the measured output signal of the NAND operation when 10 kHz input signals were applied to both gate electrodes. When neither $V_{input\_1}$ nor $V_{input\_2}$ was applied, or when only one input was applied, $V_{output}$ was measured as 1 because the resistance of the MEMS switches connected in series was significantly higher than the load resistor. Conversely, when both $V_{input\_1}$ and $V_{input\_2}$ were applied, $V_{output}$ was measured as 0 because the resistance of the MEMS switches connected in series became small compared to the load resistor. For the NOR operation, the circuit was configured by connecting a load resistor in series with MEMS switches arranged in parallel, as shown in Fig. 6d. To implement the NOR operation using the structure of the SP4T MEMS device, a circuit was configured as shown in Fig. 6e. Figure 6f shows the measured output signal of the NOR operation when 10 kHz input signals were applied to both gate electrodes at 5.8 K. When neither $V_{input\_1}$ nor $V_{input\_2}$ was applied, $V_{output}$ was measured as 1 because the resistance of the MEMS switches connected in parallel was significantly higher than the load resistor. However, if either $V_{input\_1}$ or $V_{input\_2}$ was applied, $V_{output}$ was measured as 0 because the resistance of the MEMS switches connected in parallel became negligible compared to the load resistor (Fig. 6f) . These results indicate that the MEMS switch can successfully perform logical operations even at extremely low temperatures, highlighting its potential for advanced cryogenic applications. While the MEMS switches show excellent performance at cryogenic temperatures through FEM simulations and various measurements, certain challenges remain to be addressed. Since the MEMS switch operates through the electrostatic actuation, dielectric charging occurs with repeated operation. This issue is particularly pronounced at cryogenic temperatures, where charge carriers lack sufficient thermal energy to escape the traps within the dielectric[34]. As a result, the dielectric charging phenomenon becomes more severe, potentially leading to the stiction phenomenon. Indeed, stiction is observed in some devices when they were repeatedly operated at high frequencies exceeding 100 kHz at the cryogenic temperature between 5 K to 10 K. Upon returning to room temperature, some devices resumed normal operation, indicating that the stiction was caused by dielectric charging. Given that high-speed operation is critical for quantum computing applications, further research into materials is required to mitigate dielectric charging and its effects under cryogenic conditions. Alternatively, FEM-based reliability prediction studies may also be necessary[23].

**Discussion**

This study demonstrates the suitability of commercial MEMS switch as key components for cryogenic multiplexers in large-scale quantum computing systems. The MEMS switches demonstrated stable operation at cryogenic temperature (approximately 5.8 K), exhibiting performance improvements in DC and RF characteristics compared to their room-temperature results. Furthermore, by introducing an engineered waveform, stable dynamic operation was achieved by effectively mitigating the undesired bouncing phenomenon at the cryogenic temperature. Notably, the MEMS switches performed reliably over 100 million cycles at cryogenic temperature. The successful operation of the SP4T MEMS device, along with the demonstration of logical operations such as NAND and NOR gates, underscores the suitability of MEMS switches for complex interconnect architectures in quantum systems. However, challenges such as dielectric charging-induced stiction at high frequencies persist, necessitating further advancements in materials and design optimization for high-speed and long-term reliability. Our findings support the MEMS switches as a promising candidate solution for scalable, reliable quantum systems, paving the way for the realization of large-scale quantum computing with millions of qubits.

**Materials and methods**

**FEM simulation.** Finite Element Method (FEM) simulations were conducted using ANSYS Mechanical to analyze the mechanical stress, beam deflection, and pull-in voltage of the MEMS switch under varying temperature conditions.

**Electrical measurements.** All electrical measurements were conducted using a cryogenic vacuum probe station (Lake Shore Cryotronics, CRX-VF). To determine the pull-in voltage and on-resistance between the top and bottom electrodes, a sourcemeter (Keithley 2450) was employed, configured with a reading voltage of 1 V and a compliance current limit of 10 mA. A network analyzer (Agilent N5225A) was used to characterize the insertion loss and isolation properties of the MEMS switches. For dynamic response measurements, a gate pulse generated by a function generator (Agilent 33220A) was amplified using a voltage amplifier (Tegam 2450 Precision Power Amplifier) and subsequently applied to the gate electrode of the MEMS switch. The resulting output signal was monitored and recorded using an oscilloscope (Tektronix 4 Series B MSO Mixed Signal Oscilloscope).

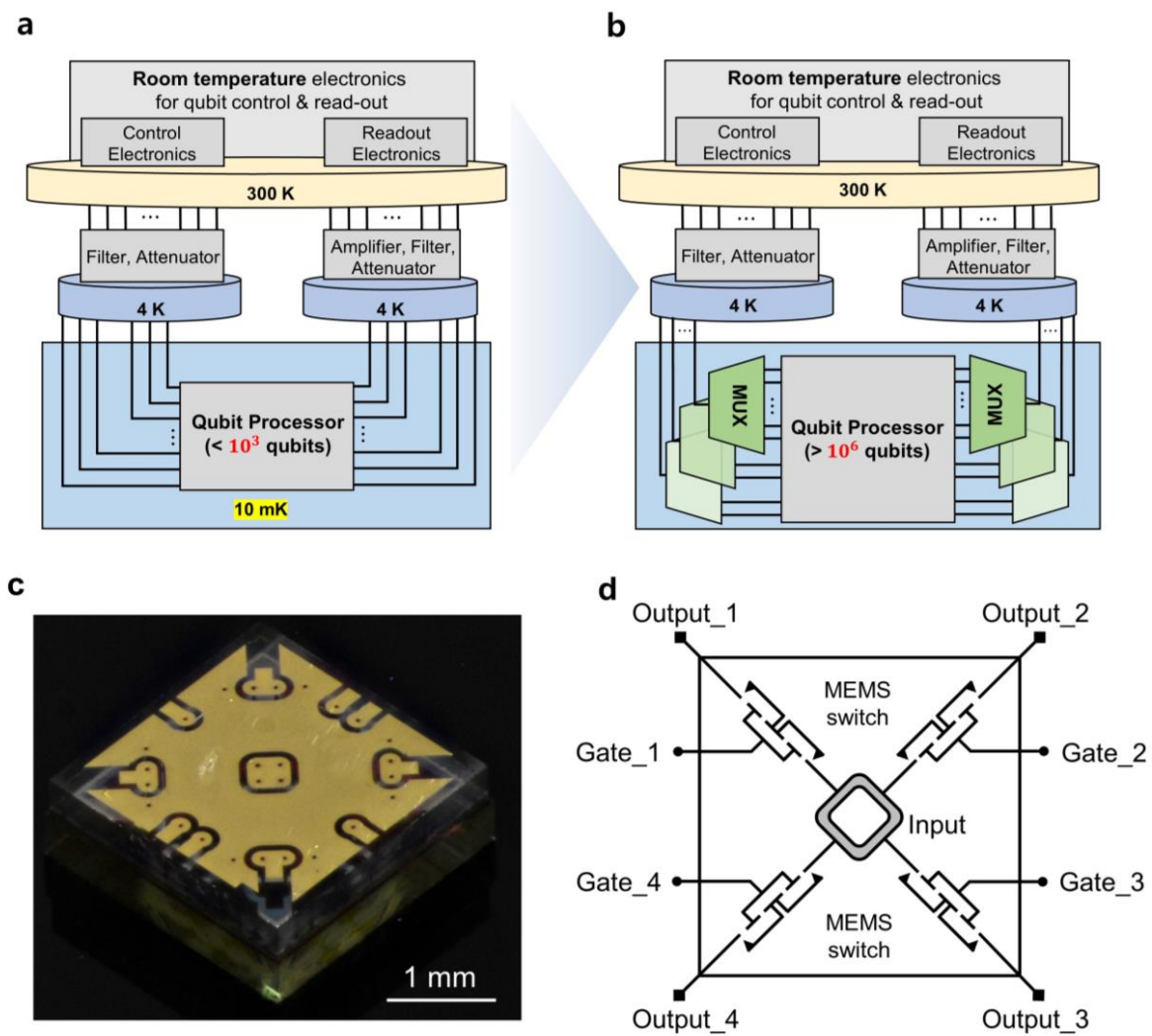

**Fig. 1 | Commercial microelectromechanical system (MEMS) switches for large-scale quantum computing systems. a** Layered architectures of current superconducting quantum computing systems. **b** Future superconducting quantum computing systems with cryogenic multiplexers. **c** Optical photograph of the SP4T device comprising MEMS switches developed by Menlo Microsystems, Inc. **d** Block diagram of the SP4T MEMS device.

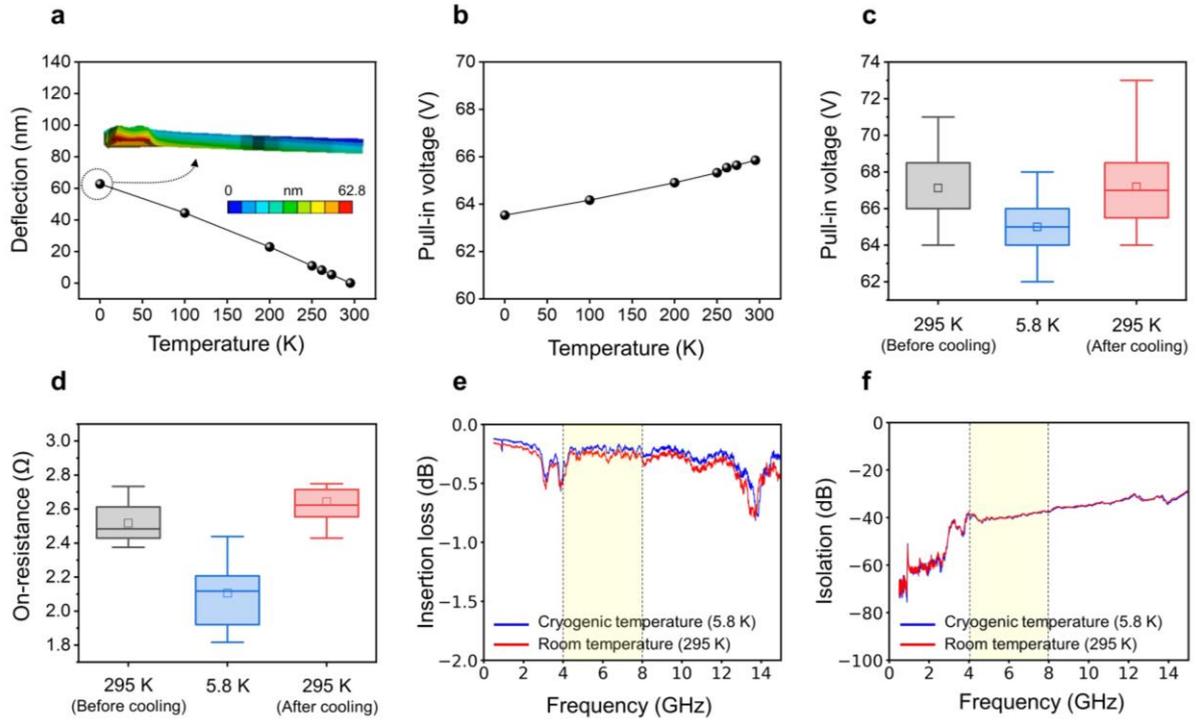

**Fig. 2 | Performance evaluation of MEMS switches at cryogenic temperature. a** Simulated deflection as a function of the temperature. **b** Simulated pull-in voltage as a function of the temperature. **c** Pull-in voltage measured sequentially at room temperature (left), at cryogenic temperature (middle), and room temperature again (right). **d** On-resistance measured sequentially at room temperature (left), at cryogenic temperature (middle), and room temperature again (right). **e** Measured insertion loss at both room and cryogenic temperatures. The insertion loss remains below 0.5 dB within the highlighted qubit frequency range of 4–8 GHz. **f** Measured isolation at both room and cryogenic temperatures.

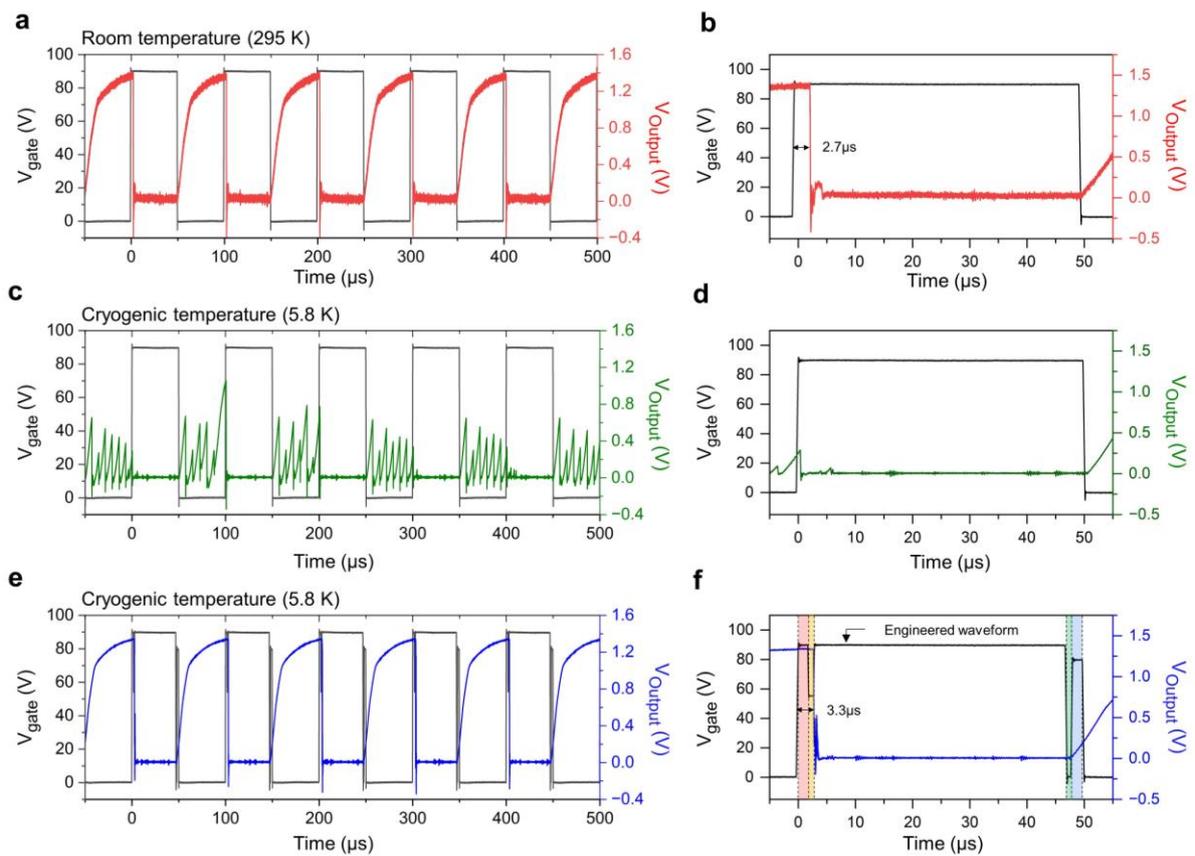

**Fig. 3 | Dynamic response of MEMS switches. a-b** Measured dynamic response and magnified view at room temperature. **c-d** Measured dynamic response and magnified view at cryogenic temperature, showing a severe bouncing phenomenon. **e-f** Measured dynamic response and magnified view at cryogenic temperature when the engineered gate waveform was introduced.

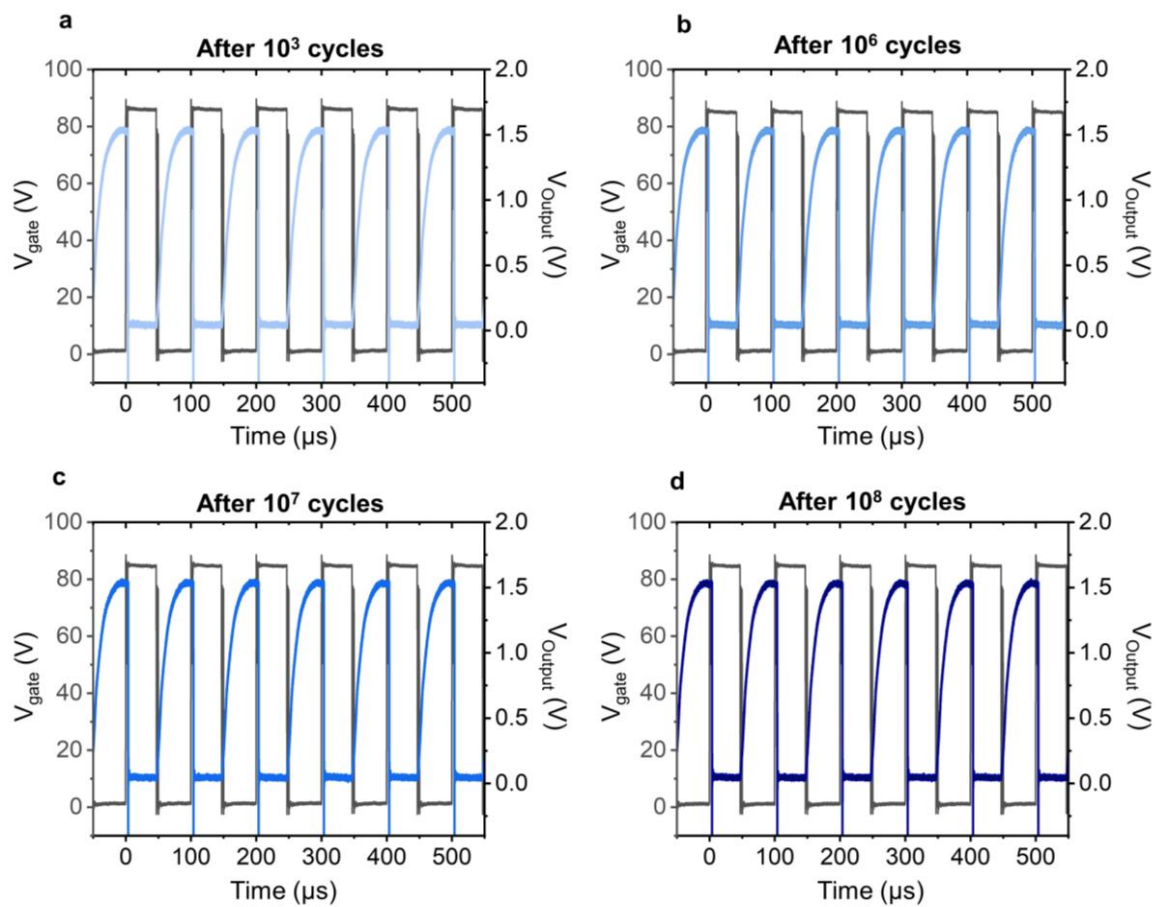

**Fig. 4 | Repetitive operation test at cryogenic temperature. a-d** Measured dynamic response results at a gate frequency of 10 kHz after $10^3$, $10^6$, $10^7$, and $10^8$ cycles, respectively.

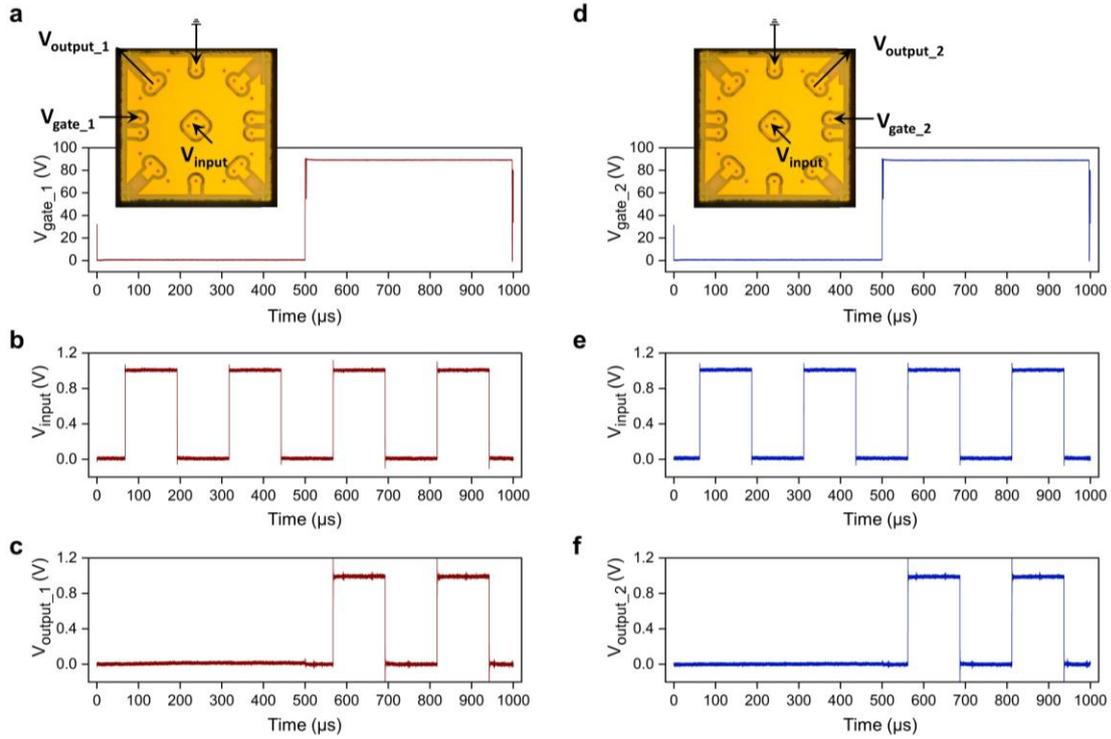

**Fig. 5 | SP4T MEMS device operation at cryogenic temperatures. a** Applied gate signal ($V_{gate\_1}$) to the SP4T MEMS device. **b** Applied input signal ($V_{input}$) to the center input electrode. **c** Measured output signal ($V_{output\_1}$) as function of the applied gate signal ($V_{gate\_1}$). **d** Applied gate signal ($V_{gate\_2}$) to the SP4T MEMS device. **e** Applied input signal ($V_{input}$) to the center input electrode. **f** Measured output signal ($V_{output\_2}$) as function of the applied gate signal ($V_{gate\_2}$).

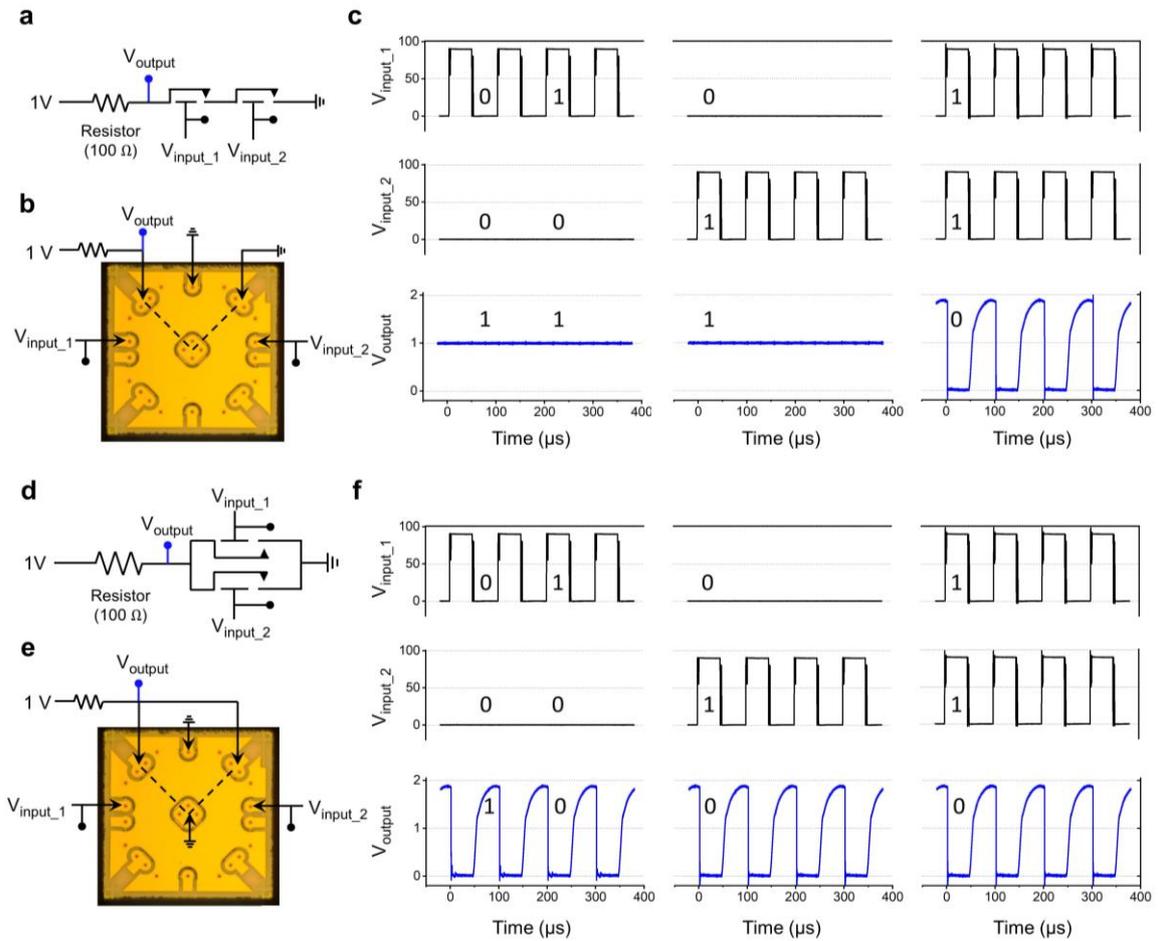

**Fig. 6 | Logical operations using SP4T MEMS device at cryogenic temperature. a** Voltage divider circuit diagram utilizing MEMS switches for NAND operation. **b** Circuit configuration of the SP4T MEMS device for NAND operation. **c** Measured output response ($V_{ouput}$) as a function of input voltages ($V_{input\_1}$ and $V_{input\_2}$). **d** Voltage divider circuit diagram utilizing MEMS switches for NOR operation. **e** Circuit configuration of the SP4T MEMS device for NOR operation. **f** Measured output response ($V_{ouput}$) as a function of input voltages ($V_{input\_1}$ and $V_{input\_2}$)

**Acknowledgements**

This research was developed with funding from the Asian Office of Aerospace Research and Development (AOARD) under the award FA2386-21-1-4088. The views, opinions and/or findings expressed are those of the authors and should not be interpreted as representing the official views or policies of the Department of Defense or the U.S. Government. This manuscript is approved for public release; distribution A: distribution unlimited.

This work is also supported by funding from Menlo Microsystems, Inc.

**Conflict of interest**

The authors declare no competing interests.

**Author contributions**

Y-B.L. performed experimental work and data analysis. C.D. contributed to data analysis and assisted with cryogenic measurement. X.Z. and N.Y. were responsible for device design and developing the test standard. Y.G. performed FEM simulation. S.A.B. supervised the research and provided critical guidance throughout the study. All the authors discussed the results and contributed to the manuscript.

# Supporting Information for

# Cryogenic Performance Evaluation of Commercial SP4T Microelectromechanical Switch for Quantum Computing Applications

*Yong-Bok Lee[1], Connor Devitt[2], Xu Zhu[3], Nicholas Yost[3], Yabei Gu[3], Sunil A. Bhave[2]\**

[1]Department of Electronics & Computer Engineering, Chonnam National University, 77 Yongbong-ro, Buk-gu, Gwangju 61186, Korea

[2]OxideMEMSS Lab, Elmore Family School of Electrical and Computer Engineering, Purdue University, West Lafayette, IN 47907 USA

[3]Menlo Microsystems, Inc., Albany, NY 12203 USA.

\*Correspondence to: bhave@purdue.edu

# Table of contents

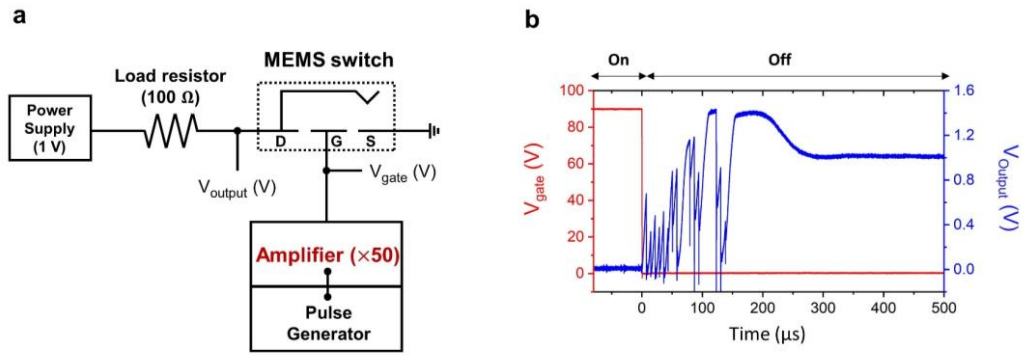

**Figure S1. Measurement of the dynamic response of the MEMS switch at cryogenic temperature. a** The voltage divider circuit consists of a load resistor and the MEMS switch connected in series. The gate pulse from the pulse generator is amplified by a voltage amplifier before being applied to the gate electrode of the MEMS switch. **b** The dynamic response measured by voltage divider circuit shows a pronounced bouncing phenomenon, attributed to the phase transition of the gas inside the packaging.

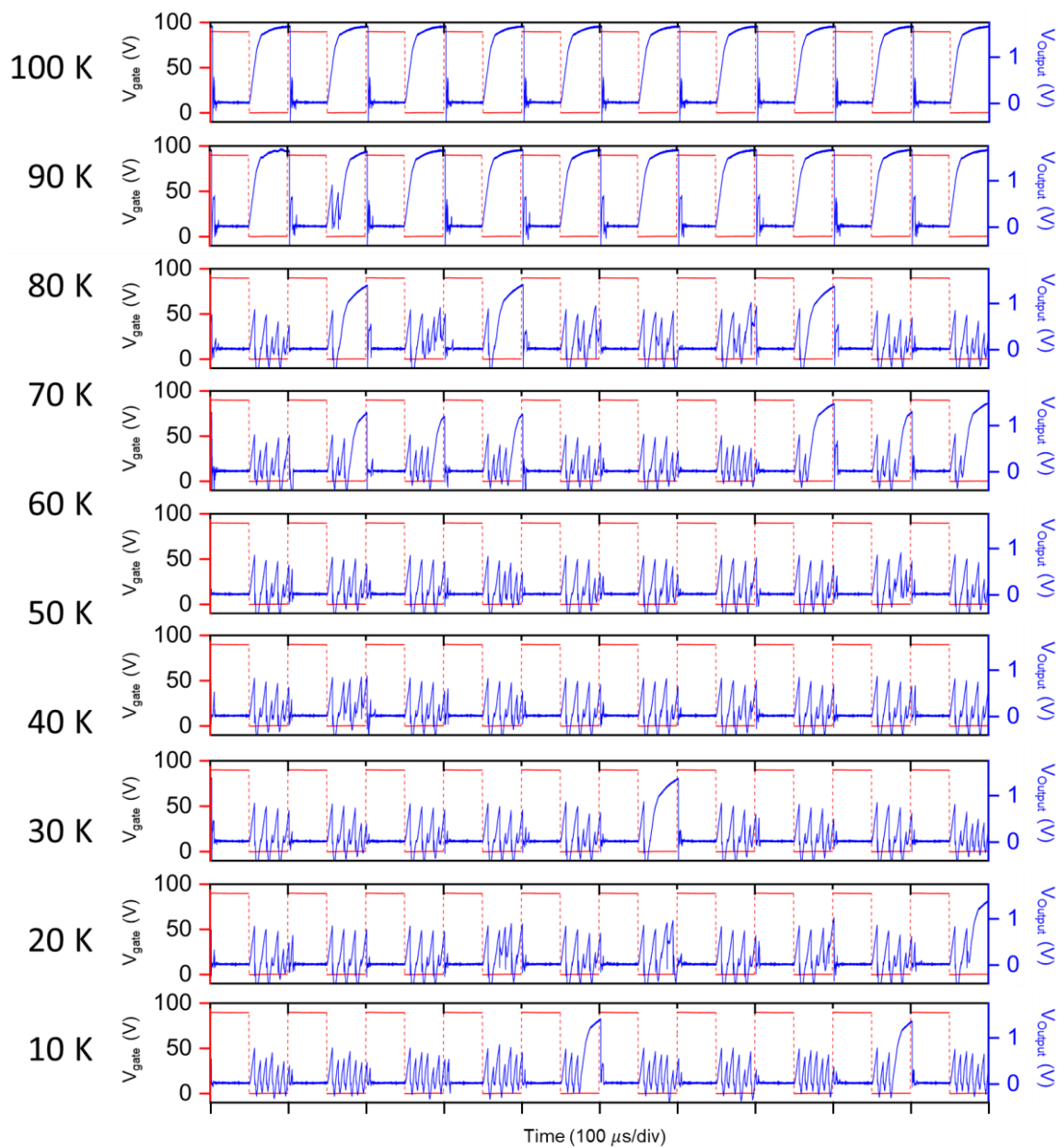

**Figure S2. Dynamic response as a function of decreasing temperature**. No bouncing phenomenon was observed above the boiling points of oxygen and nitrogen. However, as the temperature dropped below these thresholds, pronounced bouncing occurred, suggesting that the behavior was induced by the phase transition of gases sealed within the packaging.

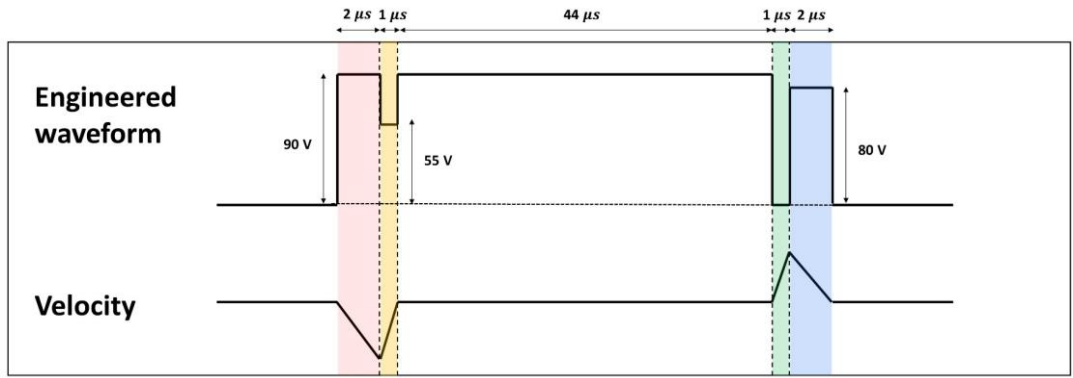

**Figure S3. Engineered waveform and the corresponding beam velocity**

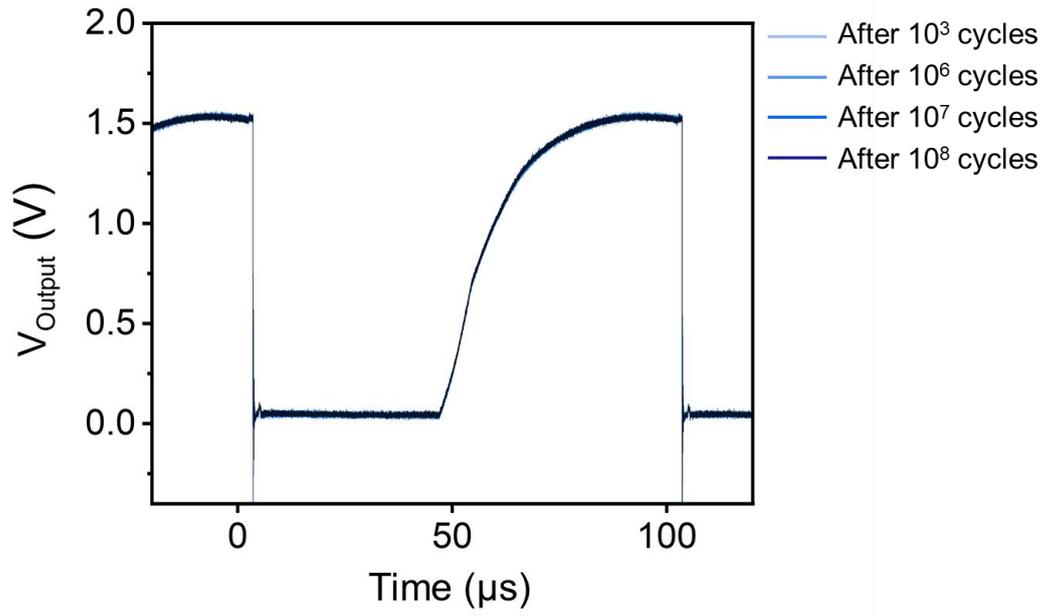

**Figure S4. Comparison of dynamic responses after $10^3$, $10^6$, $10^7$, and $10^8$ cycles at cryogenic temperature.**

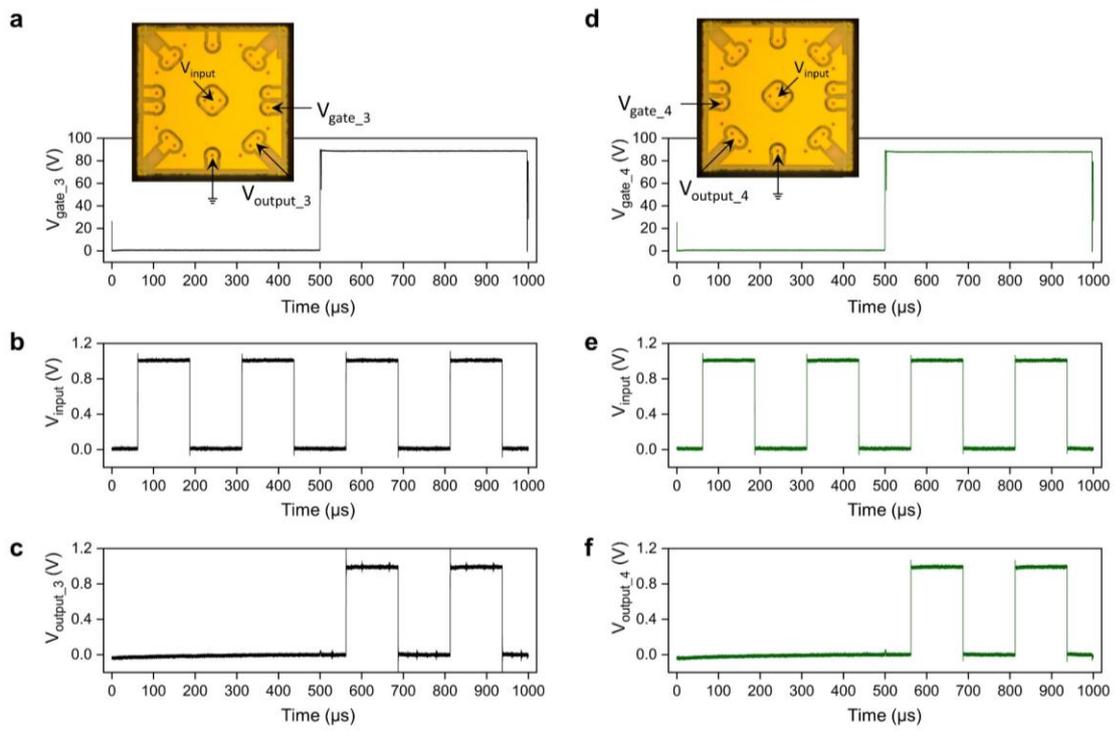

**Figure S5. Performance of the SP4T MEMS device at cryogenic temperature.** a-c Measurement results for routing the input signal to *output_3* when the operating voltage is applied to the *gate_3* electrode. d-f Measurement results for routing the input signal to *output_4* when the operating voltage is applied to the *gate_4* electrode.

7